\documentclass[journal]{IEEEtran}
\usepackage{cite}

\ifCLASSINFOpdf
\else
\fi
\usepackage[cmex10]{amsmath}
\usepackage{amssymb}

\usepackage{graphicx}

\def\be{\begin{equation}}  
\def\ee{\end{equation}}  
\def\ba{\begin{eqnarray}}  
\def\ea{\end{eqnarray}}  
\def\bc{\begin{center}}  
\def\ec{\end{center}}  
\def\p{\partial}  

\begin{document}
%
\title{Graphene-based voltage-tunable coherent terahertz emitter}
%
%
%

\author{Sergey A. Mikhailov \\ Institute 
of Physics, University of Augsburg, 
D-86135 Augsburg, Germany \\ e-mail: sergey.mikhailov@physik.uni-augsburg.de 
}

\maketitle

\begin{abstract}
\boldmath
A portion of the electromagnetic wave spectrum between $\sim 0.1$ and $\sim 10$ terahertz (THz) suffers from the lack of powerful, effective, easy-to-use and inexpensive emitters, detectors and mixers. We propose a multilayer graphene -- boron-nitride heterostructure which is able to emit radiation in the frequency range $\sim 0.1-30$ THz with the power density up to $\sim 0.5$ W/cm$^2$ at room temperature. The proposed device is extremely thin, light, flexible, almost invisible and may completely cover the needs of science and technology in the sources of terahertz radiation.
\end{abstract}


%
\IEEEpeerreviewmaketitle


\IEEEPARstart{I}{f} a fast electron moves in a periodic potential $U(x)\sim 
\sin (2\pi x/a_x)$ with the average velocity $v_0$, 
its momentum $p_x$, as well as velocity $v_x=p_x/m$, oscillates in time with the frequency 
\be 
f=\frac{v_0}{a_x}.\label{fundamfreq}
\ee
Since electrons are charged particles, such a motion is accompanied by an electromagnetic radiation with the frequency (\ref{fundamfreq}), Ref. \cite{Smith53}. 
This physical principle is used in backward-wave oscillators and free-electron lasers, where the potential $U(x)$ is produced by periodic in space electric or magnetic fields. The frequency of radiation can be tuned in these devices by varying the accelerating voltage which determines the electron velocity $v_0$. 

It was a long dream of scientists and engineers to create a compact solid-state emitter operating on the same physical principle. For example, \cite{Tsui80,Hopfel82,Okisu86,Hirakawa95} such an idea could be realized in semiconductor structures with a two-dimensional (2D) electron gas: placing a metallic grating in the vicinity of the 2D conducting layer and driving electrons across the grating stripes it seemed to be possible to force all electrons to emit electromagnetic waves at the frequency (\ref{fundamfreq}). However, instead of the strong coherent emission at the frequency (\ref{fundamfreq}) a weak thermal radiation at the frequency of two-dimensional plasmons
\be 
f_p=\sqrt{\frac{ n_se^2}{m^\star \epsilon a_x}}\label{plasmafreq}
\ee
was observed \cite{Tsui80,Hopfel82,Okisu86,Hirakawa95}; here $e$, $m^\star$ and $n_s$ are the charge, the effective mass and the surface density of 2D electrons and $\epsilon$ the dielectric constant of surrounding medium. 

The reason of this failure was explained in Ref. \cite{Mikhailov98c}. It was shown that the single-particle formula (\ref{fundamfreq}) is valid only at $f_p\ll v_0/a_x$, i.e. if the density of electrons is low or the drift velocity $v_0$ is sufficiently high. This condition is easily satisfied in vacuum devices and free-electron lasers. In a dense solid-state plasma one should take into account electron-electron interaction effects. Then one finds \cite{Mikhailov98c} that the strong coherent radiation is observed at the frequency 
\be 
\tilde f=\frac{v_0}{a_x}-f_p\label{fundamfreqFull}
\ee
{\em only} if the velocity $v_0$ exceeds a {\em threshold} value
\be 
v_{0}>v_{th}\simeq f_pa_x=\sqrt{\frac{ n_se^2a_x}{m^\star \epsilon}},
\label{threshold}
\ee
for details see \cite{Mikhailov98c}. 
Otherwise, at $v_0\ll f_pa_x$, electrons emit at the plasma frequency (\ref{plasmafreq}), just due to the heating of the system (the thermal radiation). 

The condition (\ref{threshold}) is of crucial importance for successful device operation. It shows that the surface electron density $n_s$ should be low and the drift velocity should be large. This condition is very difficult, if possible at all, to fulfil in semiconductor structures. For example, in a GaAs quantum well with $m^\star=0.067m_0$, $\epsilon=12.8$, $n_s\sim 3\times 10^{11}$ cm$^{-2}$ and $a_x\sim 1$ $\mu$m the threshold velocity (\ref{threshold}) is about $10^8$ cm/s which is more than 4 times larger than the Fermi velocity at the same density. 

\begin{figure}
\includegraphics[width=0.9\columnwidth]{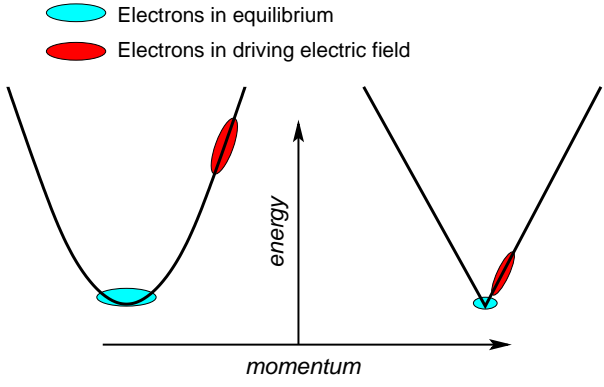}
\caption{\label{parab-lin} In order to accelerate a small amount of electrons up to velocities of order of $v_F\approx 10^8$ cm/s one needs a large change of the momentum in the case of the parabolic energy dispersion (semiconductors, left) and a much smaller change of the momentum in the case of the linear energy dispersion (graphene, right).}
\end{figure}

In graphene, a recently discovered \cite{Novoselov04,Novoselov05,Zhang05} only one-atom-thin, truly two-dimensional carbon material, the spectrum of electrons is linear, 
\be 
E_\pm ({\bf p})=\pm v_F\sqrt{p_x^2+p_y^2}\label{lin-spectrum}
\ee
with a very large, as compared to semiconductors, Fermi velocity $v_F\approx 10^8$ cm/s. This makes graphene an ideal candidate for the realization of the old idea to create a voltage tunable solid-state terahertz emitter, see illustration in Figure \ref{parab-lin}. 

However, an attempt to directly apply the above described ideas to graphene faces two difficulties. First, in graphene oscillations of the momentum do not lead to oscillations of the velocity and, hence, of the current, since the velocity ${\bf v}_\pm=\p E_\pm ({\bf p})/\p {\bf p}$ is not proportional to the momentum. Second, the condition $v_{th}<v_0\le v_F$ restricts the required electron density by the values of order of $n_s\simeq 10^{10}$ cm$^{-2}$. It is known, however, that due to internal inhomogeneities (electron-hole puddles \cite{Martin08}) the average density of electrons in a graphene sheet typically exceeds $\simeq 10^{11}$ cm$^{-2}$ even at the Dirac point.

\begin{figure}
\includegraphics[width=0.47\columnwidth]{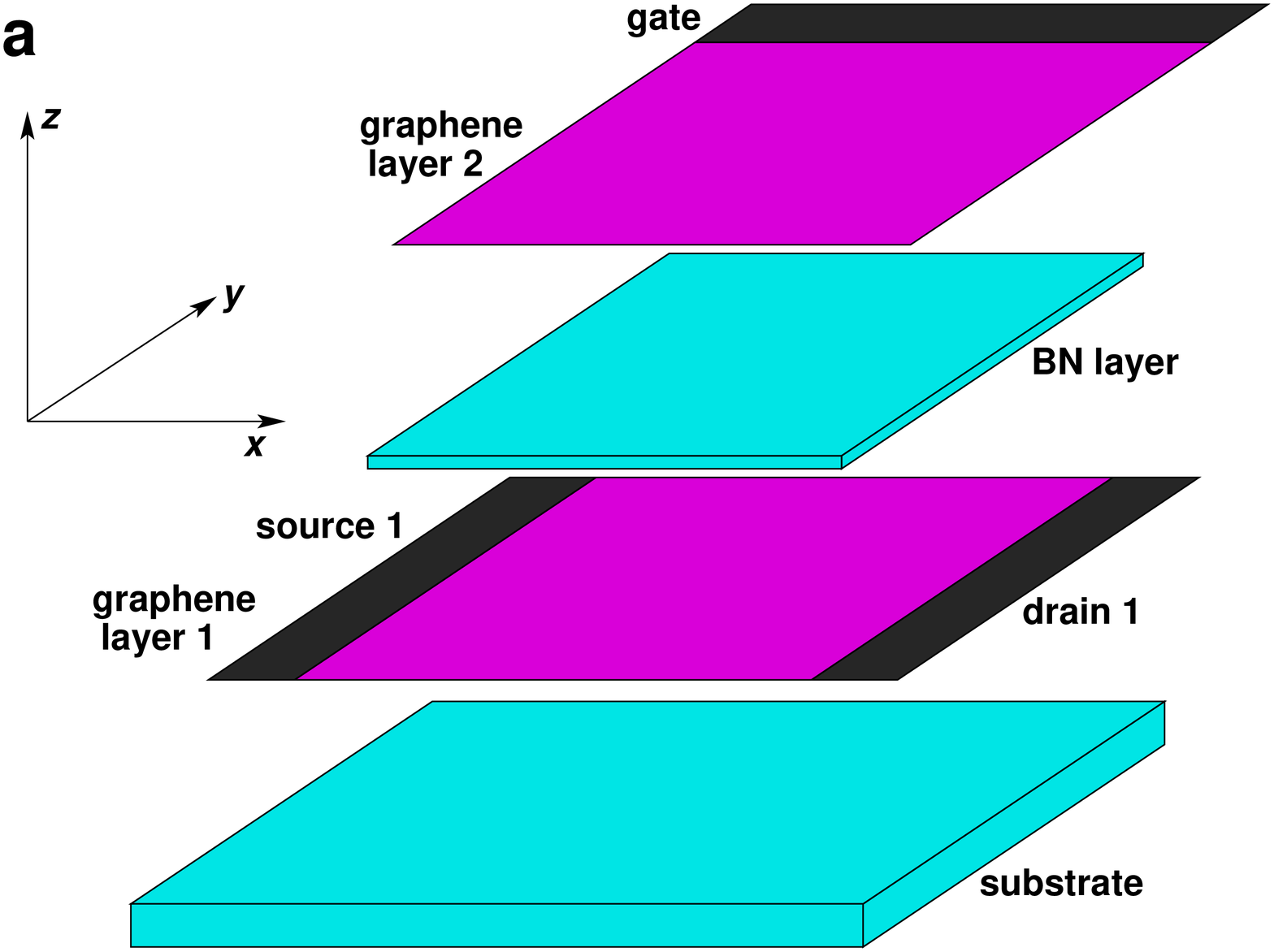}\hfill
\includegraphics[width=0.47\columnwidth]{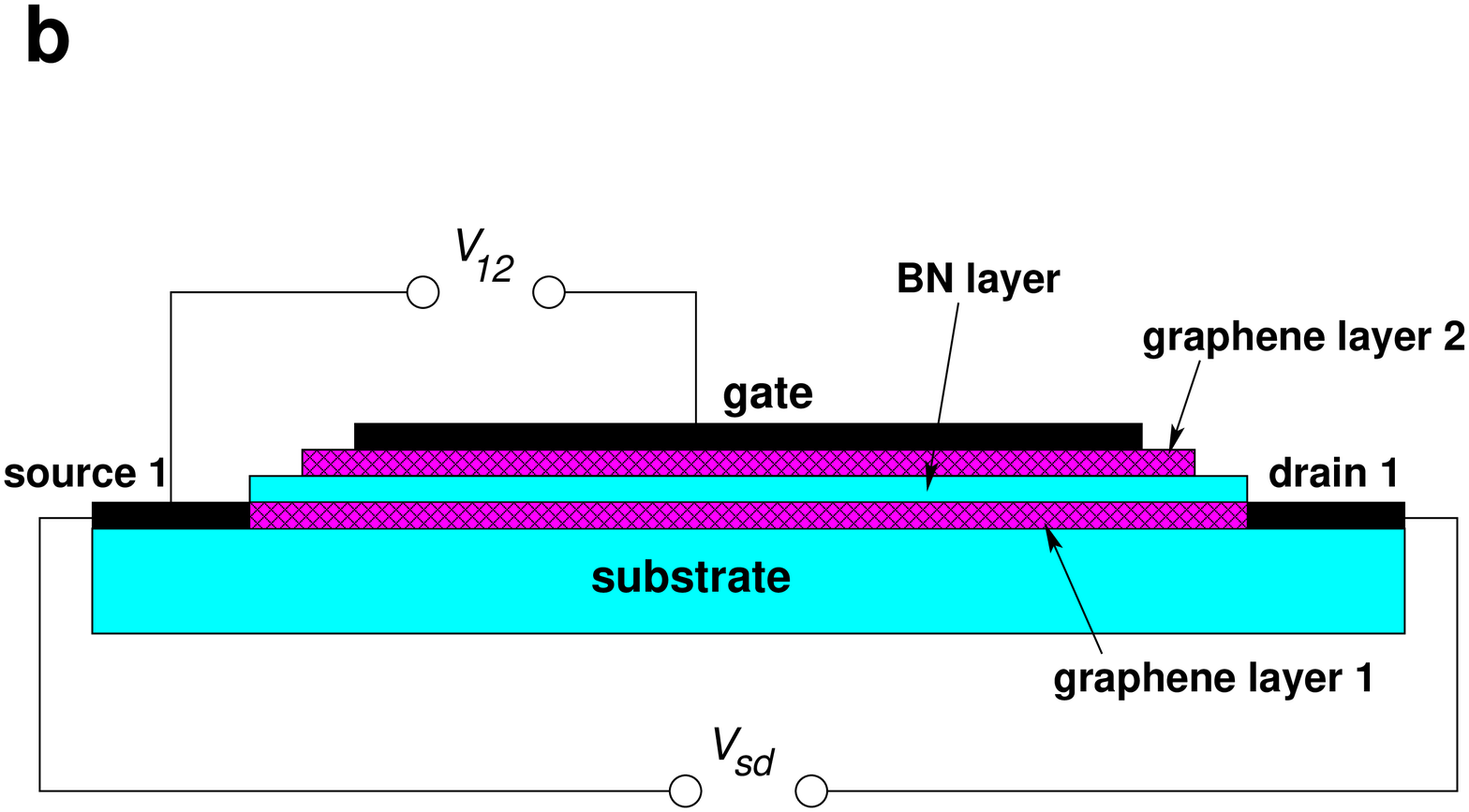} \vspace{4mm} \\
\includegraphics[width=0.47\columnwidth]{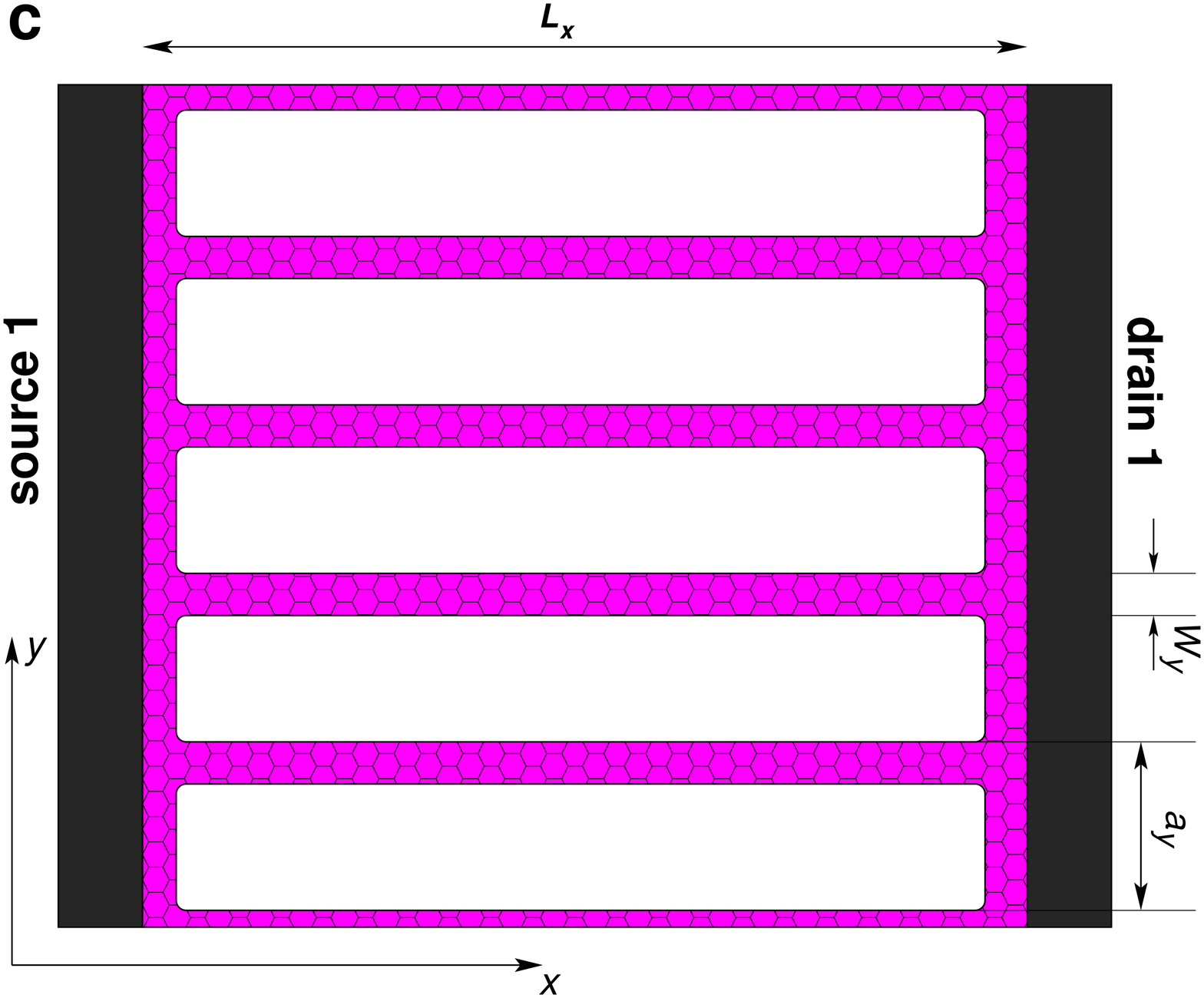}\hfill
\includegraphics[width=0.47\columnwidth]{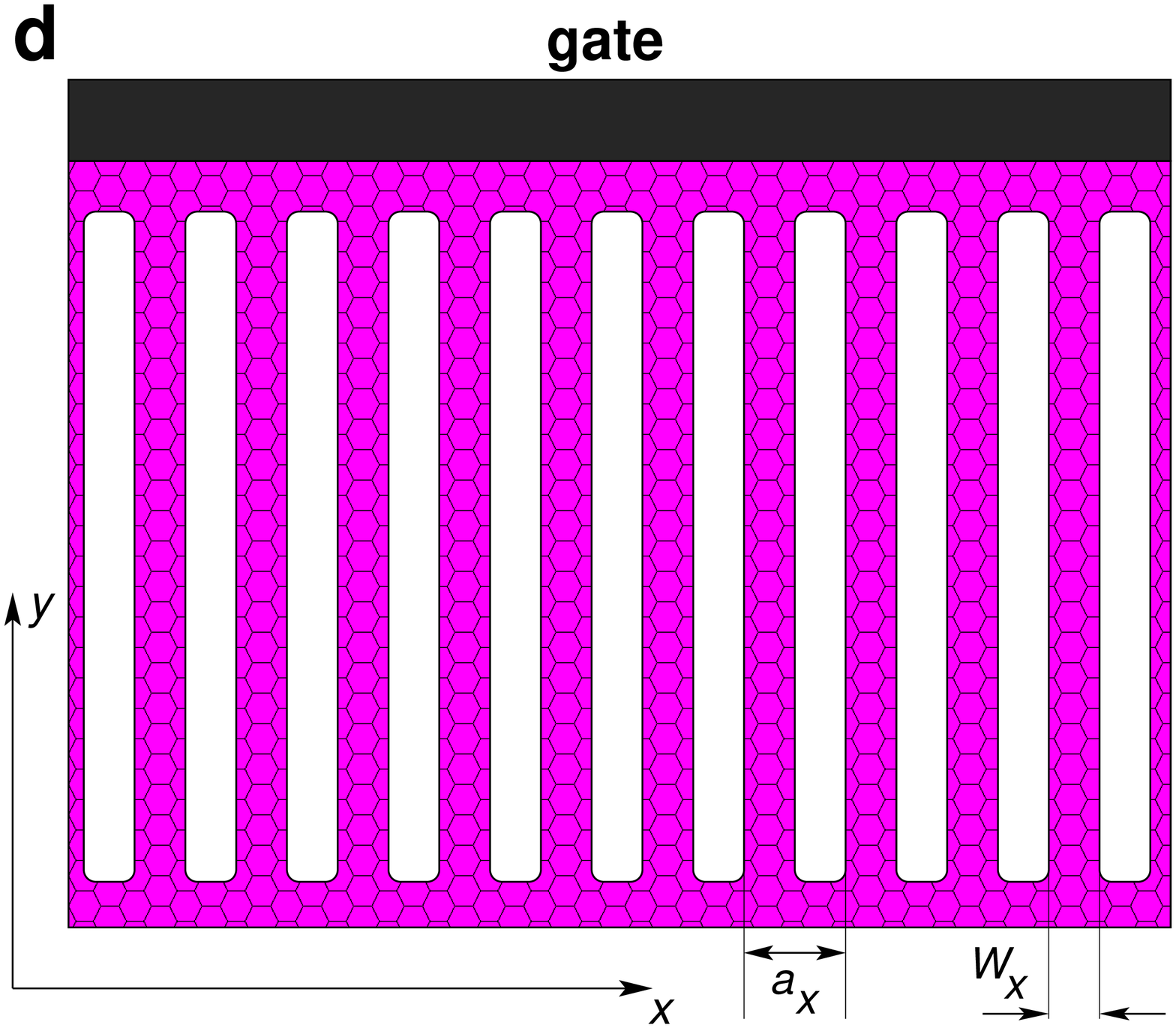} \vspace{4mm} \\
\includegraphics[width=0.47\columnwidth]{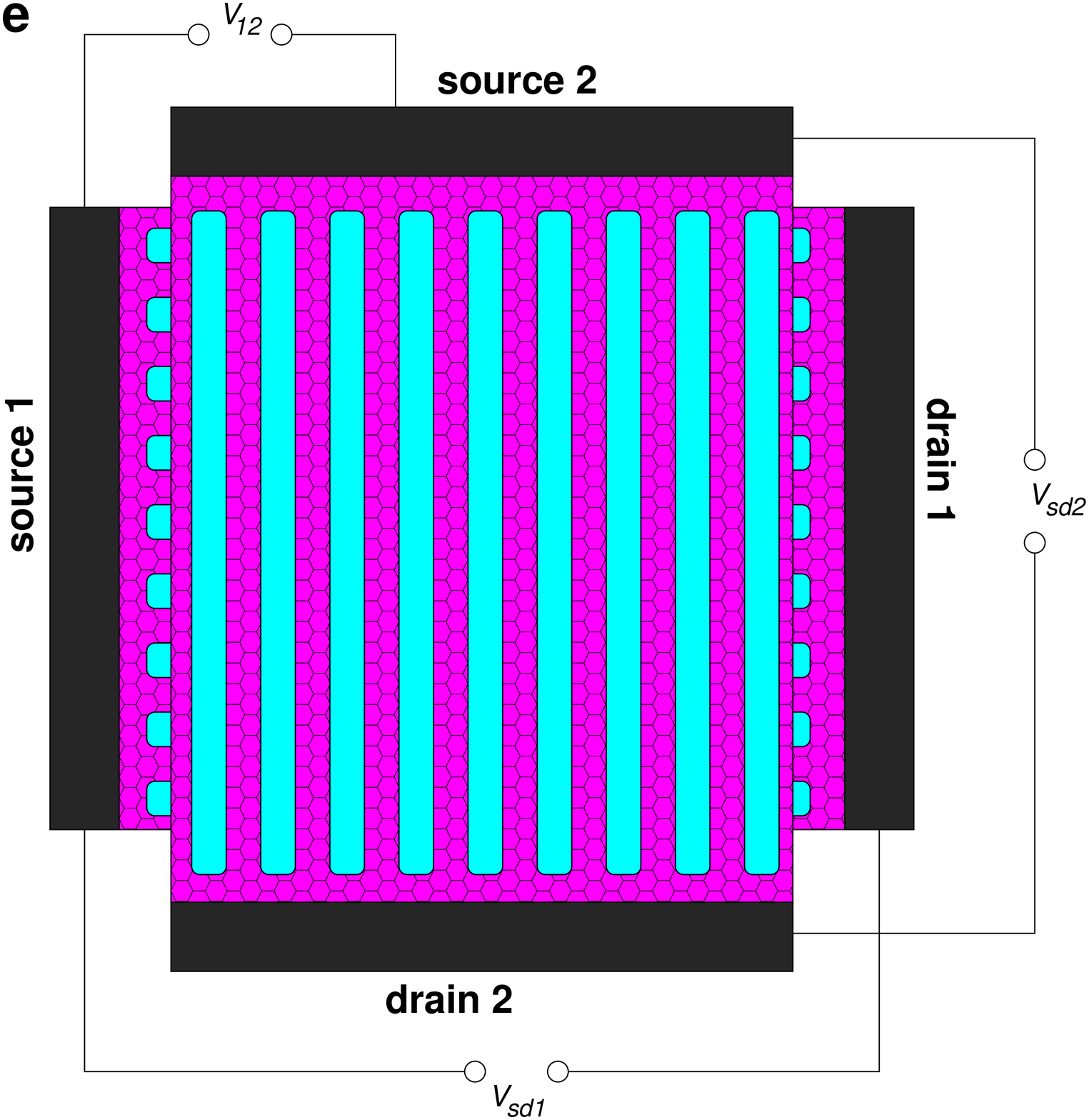}\hfill
\includegraphics[width=0.47\columnwidth]{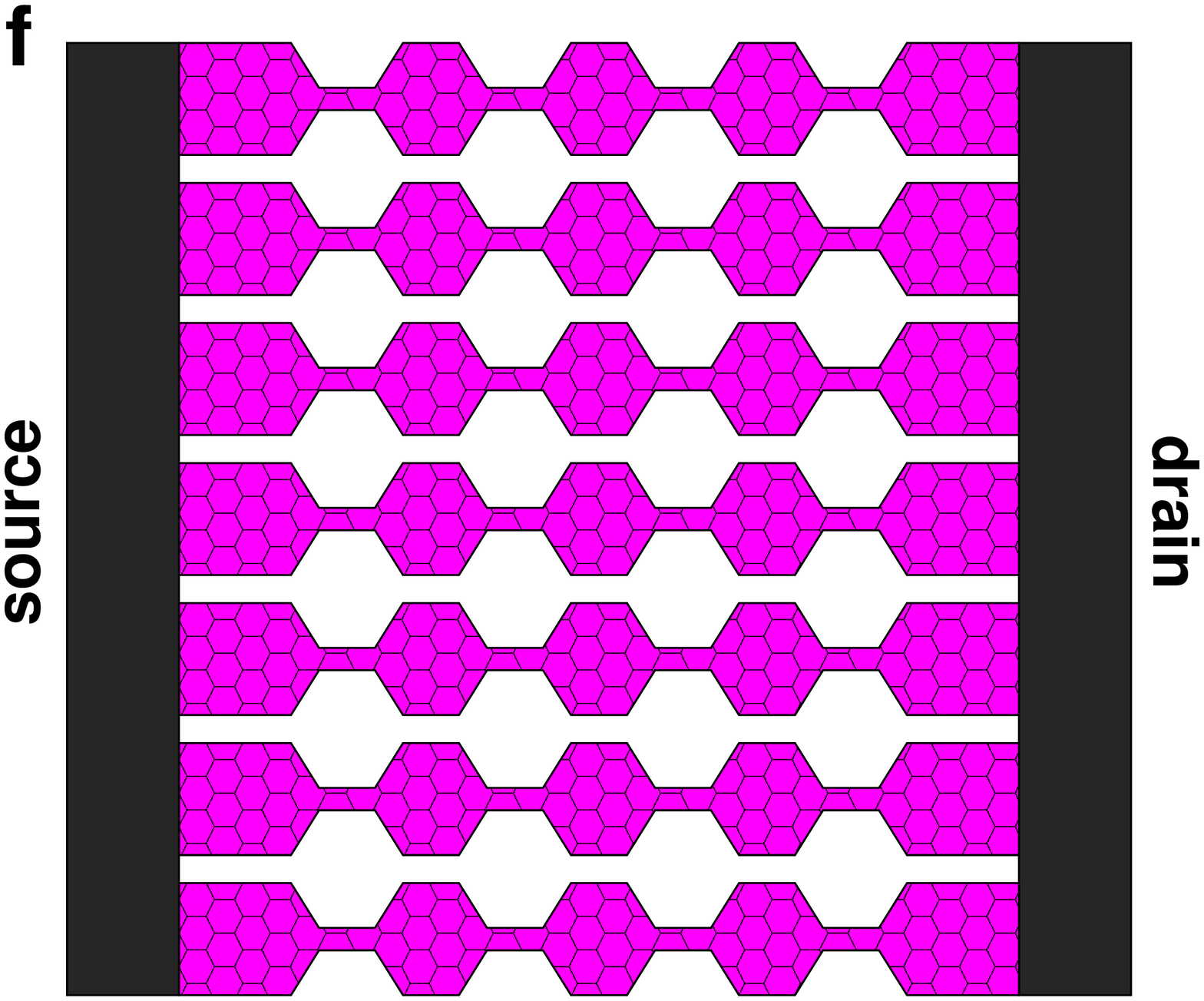} \vspace{4mm} \\
\caption{\label{genview} (a) The overall view of the device design. The first graphene layer lies on a substrate (made out of $h$-BN or SiO$_2$) and is covered by a few-nanometer thin $h$-BN dielectric layer. Two metallic contacts, ``source 1'' and ``drain 1'', are attached to the graphene layer 1 from the ``west'' and ``east'' sides. On top of the $h$-BN layer lies the second graphene layer with a metallic contact (``gate'') attached on the ``north'' (or ``south'') side. (b) The side view of the graphene based emitter. (c) The design of the 1st graphene layer. The central (operating) area of the layer is made in the form of a periodic array of narrow stripes.  (d) The design of the 2nd graphene layer (grating). (e) An alternative design of the top (grating) graphene layer with two contacts ``source 2'' and ``drain 2''. (f) A single-layer device structure with a modulated width of graphene stripes. }
\end{figure}

The both difficulties can be overcome by using, instead of continuous graphene layers, an array of narrow stripes of graphene.  This leads to the following basic design of the emitter \cite{MikhailovPatent}, Figure \ref{genview}. The first (active) graphene layer lies on a substrate made out of a dielectric material, e.g. SiO$_2$ or hexagonal boron nitride ($h$-BN), Figures \ref{genview}a. This layer consists of an array of stripes with the width $W_y$ and the period $a_y$ and has two metallic contacts ``source 1'' and ``drain 1'', Figure \ref{genview}c. Above the first graphene sheet lies a thin dielectric layer, made out of a few monolayers of $h$-BN. A second graphene layer, Figure \ref{genview}d, has the shape of a grating with the stripe width $W_x$ and the period $a_x$, oriented perpendicular to the stripes of the first layer and covers the whole structure. It has a metallic ``gate'' contact, Figure \ref{genview}d. The side view of the whole structure is shown in Figure \ref{genview}b.

\begin{figure}
\includegraphics[width=0.49\columnwidth]{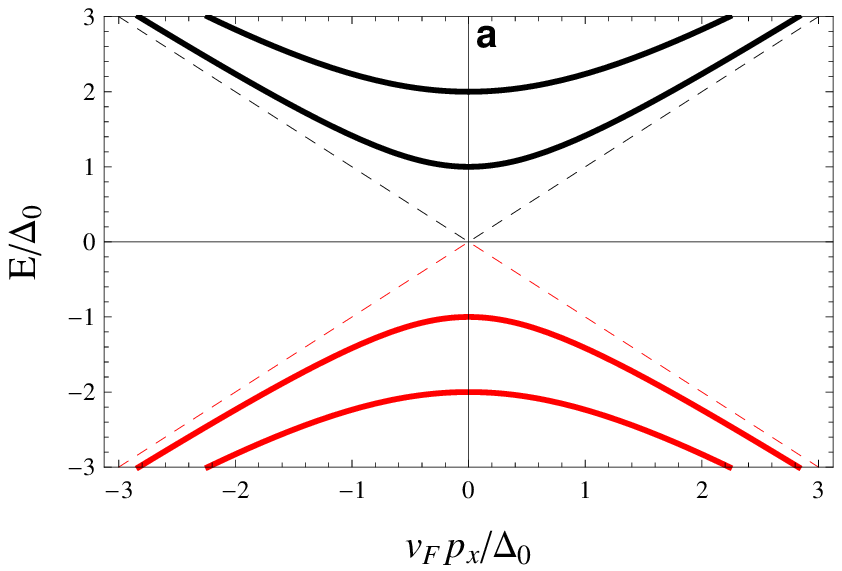}\hfill
\includegraphics[width=0.49\columnwidth]{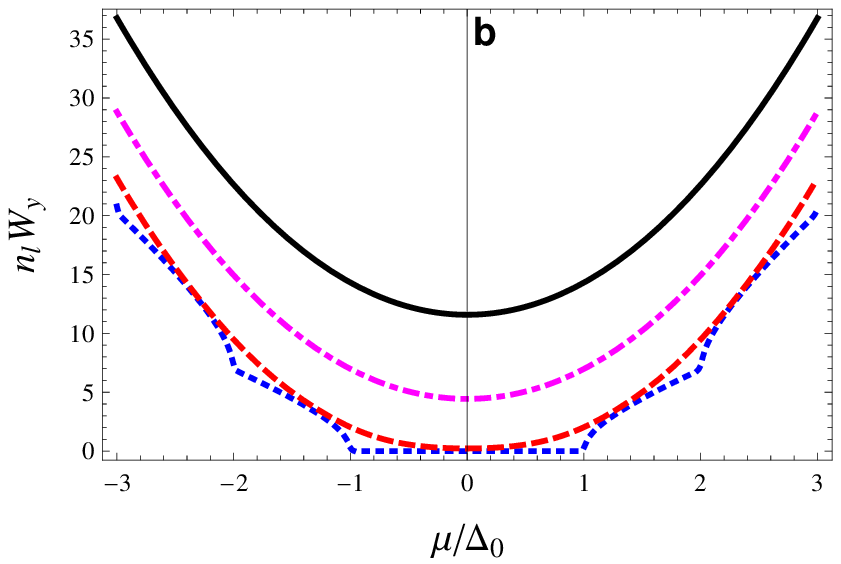}\\
\caption{\label{fig:spectr+linden} (a) The energy spectrum in the graphene stripes. If $W_y\simeq 0.1$ $\mu$m, the gap $2\Delta_0$ is about 500 K. (b) The linear charge carrier density $n_l=n_e+n_h$ as a function of the chemical potential $\mu/\Delta_0$ at different values of the temperature: $T/\Delta_0=0.01$ (dotted, blue), 0.3 (dashed, red), 0.8 (dot-dashed, magenta) and 1.2 (solid, black). If $W_y\simeq 0.1$ $\mu$m, the blue, red, magenta and black curves approximately correspond to 2.5, 75, 200 and 300 K, respectively. }
\end{figure}

Due to the finite width $W_y$ of graphene stripes in the first layer, the $p_y$-component of the momentum is quantised, $p_y\simeq \pi n/W_y$, and a gap is opened up in the graphene spectrum, 
\be 
E_{\pm,n,p_x}=\pm \sqrt{\Delta_0^2n^2+v_F^2p_x^2}
, \label{spectrum-stripe}
\ee
Figure \ref{fig:spectr+linden}a. The value of the gap, 
\be 
 2\Delta_0=2\frac{\pi\hbar v_F}{W_y},
\ee
is controlled by the stripe width $W_y$ and can be as large as $\simeq 40$ meV ($\simeq 500$ K) if $W_y\simeq 0.1$ $\mu$m. The linear density of electrons $n_e$ and holes $n_h$ per unit length then reads 
\ba 
n_e(\mu,T)&=&
\frac{g_sg_v}{W_y} \sum_{n=1}^\infty \int_{0}^\infty  \frac {dx}{1+\exp\frac{\sqrt{n^2+x^2}-\mu/\Delta_0}{T/\Delta_0}},\nonumber \\  n_h(\mu,T)&=&n_e(-\mu,T),
\label{e-density}
\ea 
where $\mu$ is the chemical potential, $T$ is the temperature and $g_s=g_v=2$ are spin and valley degeneracies. As seen from Figure \ref{fig:spectr+linden}b,  the total linear charge density $n_l=n_e+n_h$ is about $\sim 12/W_y$ at room temperature. At $W_y\simeq 0.1$ $\mu$m this gives $n_l\approx 1.2\times 10^{6}$ cm$^{-1}$. If to choose the period of the structure in the first layer $a_y$ bigger than $1.2$ $\mu$m, the average surface density $n_s=n_l/a_y$ will be smaller than $10^{10}$ cm$^{-2}$. Thus, by choosing a sufficiently large ratio $a_y/W_y$ one can always satisfy the threshold condition (\ref{threshold}). This can also be seen from the expression 
\be 
\frac{v_{th}}{v_F}=\sqrt{\frac{e^2}{\epsilon\hbar v_F}\frac{n_lW_y}{\pi}\frac{a_x}{a_y}}\label{ratio-threshold-Vfermi}
\ee
which should be smaller than unity (Eq. (\ref{ratio-threshold-Vfermi}) follows from (\ref{threshold}) if to replace $m^\star$ by $\Delta_0/v_F^2$). The first factor $e^2/\epsilon\hbar v_F$ here is of order unity. The second factor $n_lW_y$ depends on the temperature, Figure \ref{fig:spectr+linden}b, and is about 10 at room and about 1 at liquid nitrogen temperature. The last factor $a_x/a_y$ can be made as small as desired by the corresponding choice of geometrical parameters of the structure. 

In the operation mode, Figure \ref{genview}b, a large dc voltage $V_{sd}$ is applied between the source and drain contacts of the first layer and a small dc (gate) voltage $V_{12}$ -- between the first and the second graphene layer. The source--drain voltage causes a dc current in the first-layer stripes flowing in the $x$-direction. The corresponding drift velocity should lie in the window $v_{th}<v_0<v_F$. If the drift velocity is about $6\times 10^7$ cm/s and the grating period $a_x\simeq 0.2$ $\mu$m, the fundamental frequency of the emitter will be  about 3 THz. Dependent on the parameters this frequency can be red- or blue-shifted. 

Let us estimate the power of the emitted radiation. Under the action of the dc source--drain voltage electrons move with the constant drift velocity $v_0$. A small gate voltage $V_{12}$ applied between the first and the second graphene layers results in the periodic time-dependent modulation of the velocity, $v_x(t)=v_0+v_1\cos (2\pi v_0 t/a_x)$. The modulation amplitude $v_1$ is controlled by the gate voltage $V_{12}$ and can be varied between 0 and $v_0$. Assume that we have a 70\%-modulation of the drift velocity, i.e. $v_1 \simeq 4\times 10^7$ cm/s. Then the amplitude of the ac electric current density is $j_1\simeq en_sv_1$, the electric and magnetic fields of the emitted wave (calculated from Maxwell equations) are $E_1=H_1\simeq 2\pi j_1/c$ and the intensity of the emitted radiation (the Poynting vector) $W_{rad}=cE_1^2/4\pi$. At the electron density $n_s\simeq 10^{10}$ cm$^{-2}$ this gives a very large radiation power of 
\be 
W_{rad}\simeq   \frac \pi c (en_sv_1)^2\simeq 0.5 \textrm { W/cm}^2.
\label{rad-power}
\ee
Estimating the Joule's heating power 
as $W_{heat}\simeq j_{0}E_{0}\simeq j_{0}^2/\sigma_{min}$, where $\sigma_{min}=4e^2/h$ is the minimal conductivity of graphene \cite{Novoselov05}, we get 
\be 
\frac{W_{rad}}{W_{heat}}\simeq \frac{\pi j_1^2/c}{j_0^2/\sigma_{min}}\lesssim \frac{\pi\sigma_{min}}{c}=2\frac{e^2}{\hbar c},
\ee
i.e. the efficiency of the emitter is about 1\%. If the structure lies on a SiO$_2$ or BN 1 mm-thick substrate and the opposite side of the substrate is maintained at room temperature, the increase of temperature of the structure at $W_{heat}\simeq  50$ W/cm$^2$ does not exceed 10--20 K, due to the very large surface-to-volume ratio in the two-dimensional graphene and to the high thermal conductivity of the substrate. 

\begin{figure}
\includegraphics[width=0.48\columnwidth]{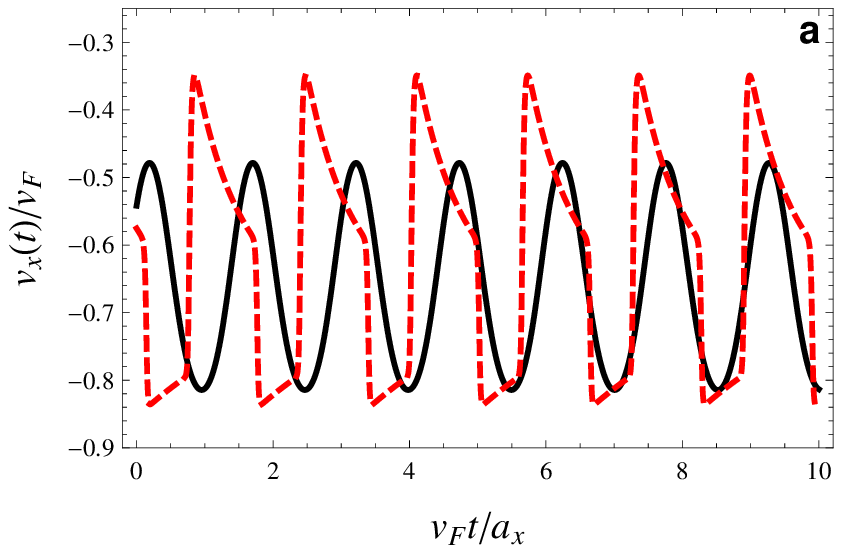}\hfill
\includegraphics[width=0.48\columnwidth]{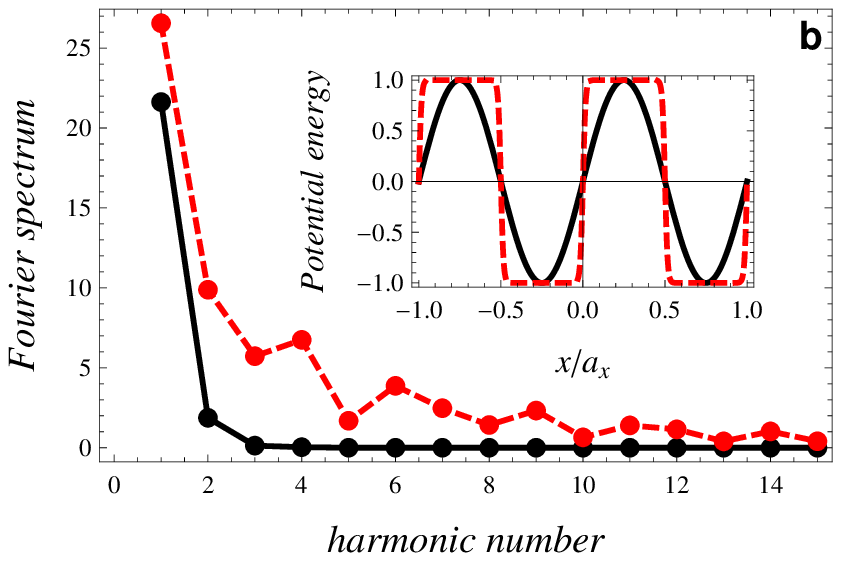}
\caption{\label{fig:velocity} (a) The time dependence of the velocity $v(t)$ calculated for $U_0/\Delta_0=0.3$, $a_x\gamma/v_F=1$, $eE_0a_x/\Delta_0=1$ and $S=0.1$ (black solid curve) and $S=10$ (red dashed curve). The average value of $v(t)/v_F$ in the cases $S=0.1$ and $S=10$ are $0.661$ and $0.615$, respectively. (b) The corresponding Fourier spectra. Inset: The model periodic potential $U(x)$ (\ref{perpot}) at $S=0.1$ (black solid curve) and $S=10$ (red dashed curve).}
\end{figure}

The proposed device may emit radiation not only at the fundamental frequency (\ref{fundamfreq}) but also at its harmonics. In the above discussion we have assumed that the periodic potential $U(x)$ produced by the voltage $V_{12}$ between the two graphene layers has a simple sinusoidal form $U(x)=U_0\sin (2\pi x/a_x)$. This is a good approximation for 
the case when the distance $D$ between the 2D electron layer and the grating is comparable with the grating period. 
In the proposed graphene--BN--graphene device the thickness of the dielectric BN layer can be as small as a few nanometers while the realistic value of the grating period is $\gtrsim 0.1$ $\mu$m. The ratio $D/a_x$ is therefore much smaller than unity and the potential $U(x)$ has a step-like form, see Inset in Figure \ref{fig:velocity}b. The higher spatial harmonics will then lead to higher frequency harmonics in the emission spectrum. To show this quantitatively, 
we solve the Newton equations of motion for an electron in the graphene stripe 
\ba 
\frac{dp_x}{dt}&=& -eE_0-\gamma p_x - \frac{dU(x)}{dx},
\label{dpdt}\\
\frac{dx}{dt}&=&v_x=\frac {v_F^2p_x}{\sqrt{\Delta_0^2+v_F^2p_x^2}},\label{dxdt}
\ea
where $E_0=V_{sd}/L_x$ is the dc electric field caused by the source-drain voltage, $L_x$ is the distance between the source and the drain and $\gamma\equiv 1/\tau$ is a phenomenological parameter describing the ``friction force'' due to the scattering. For the periodic potential $U(x)$ we use a simple model expression 
\be 
U(x)=U_0\frac{\tanh [S\sin (2\pi x/a_x)]}{\tanh S},
\label{perpot}
\ee
where $U_0$ is the potential amplitude and the parameter $S$ determines the steepness of $U(x)$, Figure \ref{fig:velocity}b: if $S\lesssim 1$, the periodic potential is smooth and close to a single-harmonic sinusoidal form; if $S\gg 1$, it has a step-like form and contains many spatial Fourier harmonics. Equations (\ref{dpdt})--(\ref{dxdt}) are written for an electron of the first subband $n=1$; evidently, taking into account the higher electron and all hole subbands will not significantly change the results. In these equations we also ignore the plasma effects since we assume that the electron density is sufficiently low, the parameter $v_{th}/v_F$, Eq. (\ref{ratio-threshold-Vfermi}), is small  and the threshold condition (\ref{threshold}) is satisfied, as has been discussed above. There are two sources of nonlinearity in equations (\ref{dpdt})--(\ref{dxdt}): the non-sinusoidal periodic potential $U(x)$ and graphene-specific nonlinear velocity-momentum relation.  

Figure \ref{fig:velocity} shows results of the numerical solution of equations (\ref{dpdt})--(\ref{dxdt}) for parameters $U_0/\Delta_0=0.3$, $eE_0a_x/\Delta_0=1$ and $a_x\gamma/v_F=1$. If $2\Delta_0\simeq 500$ K $\simeq 40$ meV 
the first two numbers correspond to $U_0\simeq 6$ meV and $E_0\simeq 1000-200$ V/cm for the grating period $a_x\simeq 0.2-1$ $\mu$m respectively. The last number means that the mean free path is of order of the grating period, i.e. $\simeq 0.2-1$ $\mu$m for the same values of $a_x$ (notice that the parameter $\omega/\gamma=\omega\tau$ is not very crucial: the device will work both at $\omega\tau>1$ and at $\omega\tau<1$). It is quite possible to realise all these requirements in experiments. Figures \ref{fig:velocity}a and \ref{fig:velocity}b show the time dependence of the velocity $v_x(t)$ and its Fourier spectra respectively; the black (solid) and the red (dashed) curves correspond to the small ($S=0.1$) and large ($S=10$) values of the steepness parameter $S$. One sees that at $S\ll 1$ the current $j_x(t)=en_sv_x(t)$ has an almost sinusoidal form with the dominating first and substantially weaker second frequency harmonics. The 
average drift velocity for the chosen parameters is $v_{0}/v_F\approx 0.661$ which corresponds to the fundamental frequency (\ref{fundamfreq}) $f_1=f\approx 0.661 v_F/a_x \simeq 0.66-3.3$ THz for $a_x\simeq 1-0.2$ $\mu$m. At $S\gg 1$ the time dependence of the current is strongly non-monochromatic with quite large second, third, fourth and even sixth frequency harmonics. The 
average drift velocity for the chosen parameters is $v_{0}/v_F\approx 0.615$ with the resulting fundamental frequency $f_1=f\approx 0.615 v_F/a_x$. This corresponds, for the same values of the grating period $a_x\simeq 1-0.2$ $\mu$m, to the first harmonic $f_1\simeq 0.6-3$ THz, second harmonic $f_2\simeq 1.2-6$ THz, fourth harmonic $f_4\simeq 2.4-12$ THz and so on. As seen from Figure \ref{fig:velocity}(b) even the ninth harmonic ($f_9$ is up to 27 THz) has the amplitude only one order of magnitude smaller than the first one. 
The emitted radiation is linearly polarised in the direction of the dc current ($x$-direction).

So far we have discussed the traditional field-effect-transistor-type design with the top graphene layer having only one (gate) contact, Figure \ref{genview}d. Alternatively, the top layer may have 
two metallic contacts, ``source 2'' and ``drain 2'', which gives the opportunity to drive a dc current in the second layer too, Figure \ref{genview}e. Then the system becomes symmetric, with the second (first) layer serving as the grating coupler for the first (second) layer and the opportunity to emit radiation with both $x$- and $y$-polarization independently controlled by the source-drain voltages $V_{sd1}$ and $V_{sd2}$. Another possible embodiment employing a single graphene layer is shown in Figure \ref{genview}f. The periodic potential $U(x)\propto 1/W_y(x)$ is produced in this case by the alternating stripe width $W_y(x)$. An advantage of this design is that it employs only one graphene layer. In this case, however, the mean free path should exceed the grating period $a_x$.

In addition to the ability to work as an emitter, the proposed device may also operate as a new type of a field-effect-transistor (amplifier) combined with a plane radiating antenna. Indeed, the intensity of the emitted radiation is determined by the amplitude of the ac electric current $j_1$. This amplitude depends, in its turn, on the voltage $V_{12}$ between the first and second graphene layers: if $V_{12}=0$, the current flowing from the source to the drain does not depend on time and no radiation is emitted; if $V_{12}\neq 0$, the current gets modulated and the device emits an electromagnetic signal. The radiation intensity can be very large ($\simeq 5$ kilowatts from square meter, Eq. (\ref{rad-power})) at the few-millivolt input signal $V_{12}$ (see estimates above). The proposed device thus amplifies an input signal and sends it directly to the surrounding space within the same physical process. 

The single-layer graphene absorbs only 2.3\% of visible light \cite{Nair08}. The $h$-BN layer is a dielectric with a large band gap and is also transparent. The proposed powerful terahertz devices will therefore be very thin, light and  practically invisible. Furthermore, such few-nm thick devices can be bent up focusing radiation and producing a huge concentration of THz power in a very small spatial volume. 

To summarise, we have proposed a new type of few-atomic-layers thin, light, bendable, almost invisible and voltage tunable graphene based device which is able to produce a powerful electromagnetic radiation in a broad range of terahertz frequencies. This device may cause a revolution in the terahertz science and technology and will have a lot of applications in many different areas.

\section*{Acknowledgment}

The author would like to thank Ulrich Eckern and Roland Grenz for interest to this work and Geoffrey Nash, Fabrizio Castellano and Jerome Faist for useful discussions. This work was supported by Deutsche Forschungsgemeinschaft.

\ifCLASSOPTIONcaptionsoff
  \newpage
\fi


\bibliographystyle{IEEEtran}
\end{document}